\documentclass[11pt,onecolumn]{article}
\usepackage[utf8]{inputenc}
\usepackage{amsmath,amsfonts,amssymb,graphicx,authblk,placeins,cite}
\usepackage{comment}
\usepackage{color,soul}
\usepackage{hyphenat}
\usepackage[top=50pt,bottom=50pt,left=70pt,right=70pt]{geometry}
\usepackage{sectsty}
\usepackage[font=sf]{caption}
\usepackage{hyperref}

\begin{document}

\allsectionsfont{\sffamily}
\pagenumbering{gobble}

\title{\sffamily \Huge Multi-Modal Cognitive Maps based on Neural Networks trained on Successor Representations}
 \vspace{1cm}

\author[1,2]{\sffamily Paul Stoewer}
\author[1,3]{\sffamily Achim Schilling}
\author[2]{\sffamily Andreas Maier}
\author[1,3,*]{\sffamily Patrick Krauss}


\affil[1]{\small{Cognitive Computational Neuroscience Group, University Erlangen-Nuremberg, Germany}}

\affil[2]{\small{Pattern Recognition Lab, University Erlangen-Nuremberg, Germany}}

\affil[3]{\small{Neuroscience Lab, University Hospital Erlangen, Germany}}

\affil[*]{\small{corresponding author}}

\maketitle


\begin{abstract}{\sffamily \noindent

Cognitive maps are a proposed concept on how the brain efficiently organizes memories and retrieves context out of them. The entorhinal-hippocampal complex is heavily involved in episodic and relational memory processing, as well as spatial navigation and is thought to built cognitive maps via place and grid cells. To make use of the promising properties of cognitive maps, we set up a multi-modal neural network using successor representations which is able to model place cell dynamics and cognitive map representations. Here, we use multi-modal inputs consisting of images and word embeddings. The network learns the similarities between novel inputs and the training database and therefore the representation of the cognitive map successfully. Subsequently, the prediction of the network can be used to infer from one modality to another with over $90\%$ accuracy. The proposed method could therefore be a building block to improve current AI systems for better understanding of the environment and the different modalities in which objects appear. The association of specific modalities with certain encounters can therefore lead to context awareness in novel situations when similar encounters with less information occur and additional information can be inferred from the learned cognitive map. Cognitive maps, as represented by the entorhinal-hippocampal complex in the brain, organize and retrieve context from memories, suggesting that large language models (LLMs) like ChatGPT could harness similar architectures to function as a high-level processing center, akin to how the hippocampus operates within the cortex hierarchy. Finally, by utilizing multi-modal inputs, LLMs can potentially bridge the gap between different forms of data (like images and words), paving the way for context-awareness and grounding of abstract concepts through learned associations, addressing the grounding problem in AI.
}
\end{abstract}

\newpage
\section*{Introduction}

Real understanding of the general environment and retrieving context from the current situation is still a challenge for current AI systems \cite{mahowald2023,marcus2018deep}. The human brain, however, can make sense of novel and complex situations even with little information present. The hippocampus plays a crucial role in that domain, since it is receiving highly processed information and uses them for memory processing \cite{opitz_memory_2014}, as well as spatial and non-spatial navigation \cite{epstein2017cognitive,park2021inferences, killian_grid_2018}. It is also part of the limbic system and therefore even receives emotional information \cite{rolls2015limbic}. Place cells \cite{o1971hippocampus} and grid cells \cite{hafting2005microstructure} have been found experimentally in the hippocampus of rodents. Whereas place cells fire at specific positions \cite{o1971hippocampus}, grid cells fire periodically distributed over the environment in a grid-like manner \cite{hafting2005microstructure}. Additionally, their pattern changes with regard to the environment \cite{derdikman_fragmentation_2009}.

\begin{figure}[htbp]
    \centering
    \includegraphics[width = 15cm]{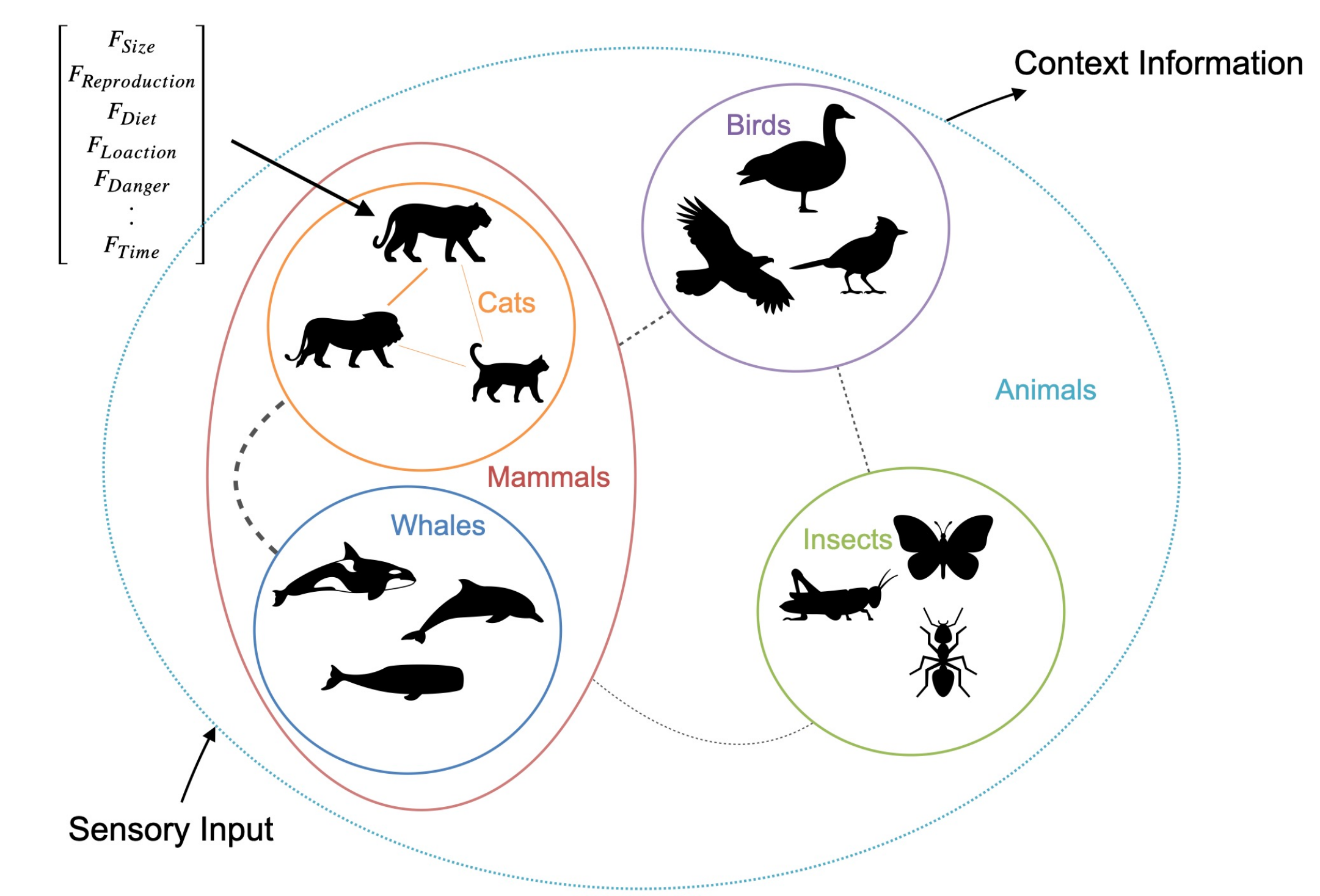}
    \caption{A schematic representation of a cognitive map with different kinds of animals. Each animal can be represented by a feature vector holding all possible information, like overall characteristics, location or time of the encounter, visual representation or any information worth storing. The objects in the map are connected to all other objects, and the weight of the connection determines the similarity of the objects. Similar objects therefore form clusters. The varying map scaling in the entorhinal-hippocampal complex \cite{collin2015memory} enables the to zoom in and out of the categories like in our examples from over all animals, to mammals to cats. A cognitive map can give therefore context information for novel inputs, by providing information stored in the map of similar encounters in the past.}
    \label{fig1}
\end{figure}

It has been proposed that the entorhinal-hippocampal complex is forming so-called cognitive maps with the help of place and gird cells \cite{o1978hippocampus, moser2017spatial}. Cognitive maps can display arbitrary kinds of information and their relationship to each other (cf. fig \ref{fig1}). These maps help to connect past experiences and combine them together in a multi-domain map, supporting flexible and higher cognition \cite{bellmund2018navigating}. For example, when encountering a lion in the jungle, the situation would be considered dangerous. In contrast, seeing a lion in the zoo would be considered safe. Ths, cognitive maps help to make quick assessments of situations when certain features are present or not, leading to more context awareness. 

There are several attempts and functional models which try to find a model description of the representations and computations in the entorhinal-hippocampal complex \cite{whittington2020tolman,stachenfeld2014design, stachenfeld2017hippocampus}. Successor representations are a mathematical framework, which was proposed to model place cell activity in the hippocampus \cite{stachenfeld2017hippocampus}.

In previous works, we were able to recreated place cell fire patters of rodent experiments, and demonstrated neural networks' ability to learn successor representations in spatial exploration and linguistic structure inference \cite{stoewer_neural_2022}. Furthermore, we explored cognitive maps of semantic spaces. In particular, we showed how maps of animal characteristics naturally emerge using handcrafted feature vectors, and how these maps enable to interpolate and restore sparse, disturbed or even unknown input feature vectors \cite{stoewer2023neural}. Simultaneously, Surendra et al. 2023 offered insights into the emergence of abstract linguistic categories within neural networks trained on a word successor task \cite{surendra2023word}. Finally, we further advanced these concepts, showcasing a model that employs successor representations, neural networks, and word embeddings to construct multi-scaled cognitive maps of semantic features clustering in three categories \cite{stoewer2023conceptual}. 

Here, we further develop this model to make use of the versatile functions of the hippocampus. In particular, we now combine different modalities in a single cognitive map. In our experiment, images simulate visual representations and are combined with word embeddings which represent the semantic features of the input. The aim of our study is to use the pre-trained cognitive map to infer the features of one modality from the other modality.

\section*{Methods}

\subsection*{Successor Representation}

To be able to build cognitive maps, we used the mathematical concept called the  successor representation (SR). The SR was firstly connected with hippocampal place cells by Stachenfeld et al, who modeled place cells with the formula \cite{stachenfeld2017hippocampus}. The SR was originally designed to build a representation of all possible future rewards $V(s)$ that may be achieved from each state $s$ within the state space over time \cite{SR_Original}. The SR $V(s)$ is calculated by including all states of the set up environment. The parameter $t$ expresses how many steps in the environment are taken from the starting state $s$ and the term $R(s_t)$ tells the Reward at each position $t$. The discount factor $\gamma[0,1]$ scales the values of future states $s_t$ that are further in the future relative to the respective initial state $s_0$ (cf. eq. \ref{successor_representation}).  

\begin{equation}\label{successor_representation}
\centering
V(s) = E[\sum^{\infty}_{t=0}\gamma^t R(s_t)|s_0=s] 
\end{equation}
Here, $E[\,]$ denotes the expectation value.

The SR matrix $V(s)$ can be described using the matrix $M$, which can be computed from the state transition probability matrix $T$ of the environment (cf. eq. \ref{SR_refactor}). In case of supervised learning, the environments used for our model operate without specific rewards for each state. For the calculation of these SR matrices, we set $R(s_t)=1$ for every state.

\begin{align}\label{SR_refactor}
    V(s) = \sum_{s'} M(s,s')R(s') && M = \sum^{\infty}_{t=0}\gamma^t T^t
\end{align}

\subsection*{Dataset}

For the experiment, we used the MNIST dataset for the training and test images \cite{deng2012mnist}. The training word embeddings from 0-9 were calculated via the spaCy \cite{spacy2} library. For altered test word embeddings, we added a Gaussian noise vector  with a random standard deviation ranging from $[0,0.4]$ to each embedding vector. 

\subsubsection*{Transition probability and successor representation matrix}

\begin{equation}\label{transitions}
\centering
T(s, s') = \begin{cases}
  0 & \text{if number(s) $\neq$ number(s')}, \\
  \frac{1}{n_s} & \text{if number(s) = number(s')}.
\end{cases} 
\end{equation}

The transition probability matrix of the environment is based on the similarity of the states of the environment. The transition probability from one state to another, is 0 for states which are not the same number and if the states share the same number label the average of the same number in the dataset (cf. eq. \ref{transitions}). The SR matrix is then calculated as explained before.

\subsection*{Neural network architectures and training parameters}

\begin{figure}[htbp]
    \centering
    \includegraphics[width = 10cm]{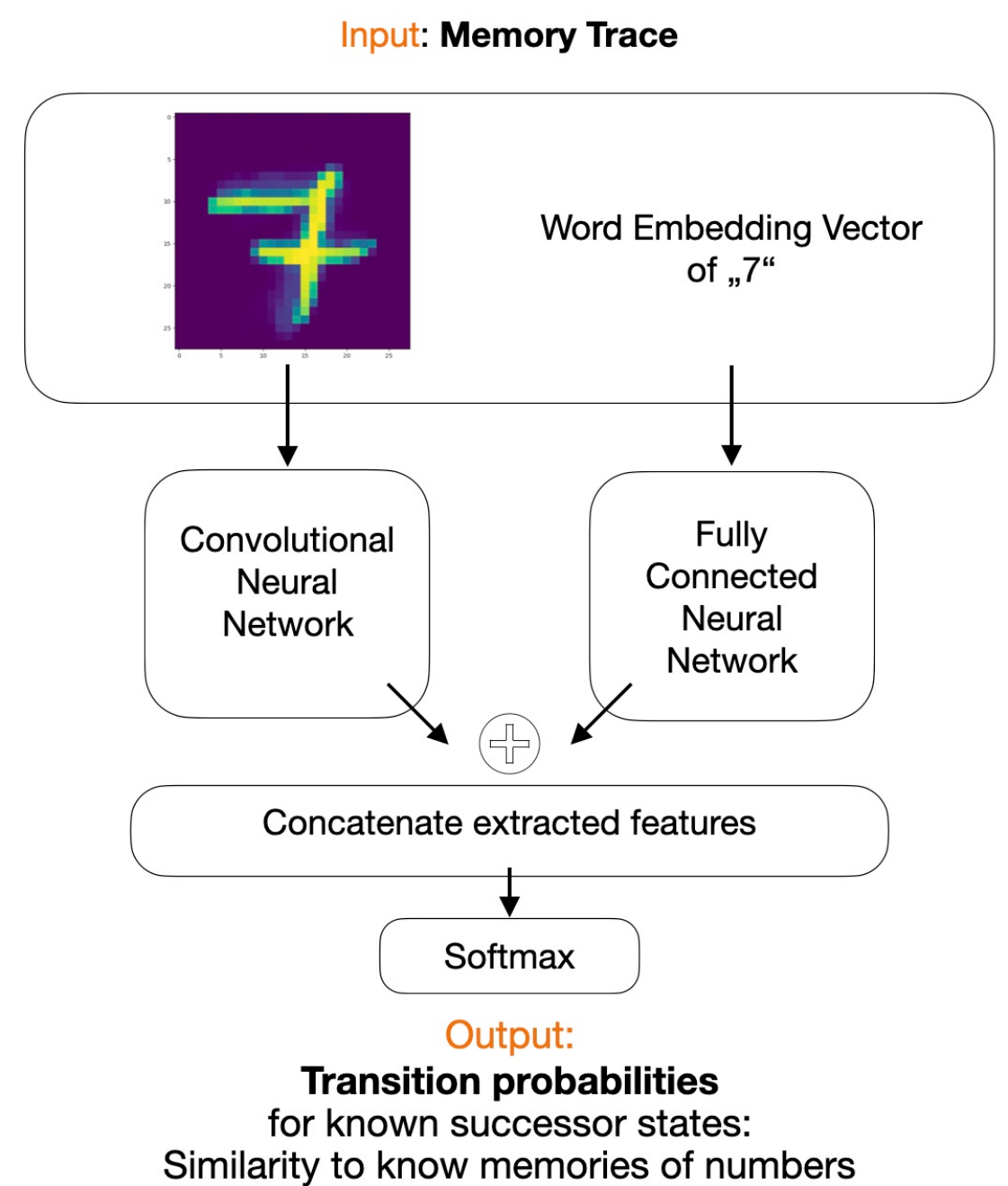}
    \caption{The neural network represented as graph diagram. The left arm of the network receives the digits images with dimensionality 24x24 as input and propagates them through a 10-layered convolutional neural network. The right arm receives word embeddings with dimensionality 300 as input, which propagate through a 5-layered fully connected multi-layer perceptron. Subsequently, the extracted features (i.e. outputs) of both arms are concatenated and serve as input for a 6-layered fully connected neural network with a final softmax layer for the output. The dimensionality of the softmax layer depends on the number of training samples used.}
    \label{fig2}
\end{figure}

To be able to process multi-modal input data, we set up two different neural networks which, concatenate their output and process them together to the final output prediction. The first network receives an image with the size of $24x24$ as input and is therefore set up as a 10 layer convolutional neural network. The second networks input is the word embedding with the size of 300 and set up as a 5 layer fully connected network. The last layer of both networks is a fully connected layer which returns the extracted features of both modalities, concatenates them and passes them into the final 6 fully connected layers and a last soft max layer for the final prediction (cf. fig \ref{fig2}). The size of the soft max layer depends on how many training samples were used for training. We trained 10 different models, with 1.000-10.000 samples and therefore the size of the layer varied in the same way. The label for the training process is the SR probability vector from the input state $s$.

To make robust predictions, the network received in $30\%$ of all cases no image, only the word embedding as input and in $35\%$ of all cases no word embedding but only the image as input. The network always received at least one of the modalities as input. All networks were trained for 1000 epochs with a batch size of 50 and an Adam optimizer with a learning rate of $1e-4$.

\subsubsection*{Interpolating unknown features}

\begin{figure}[htbp]
    \centering
    \includegraphics[width = 10cm]{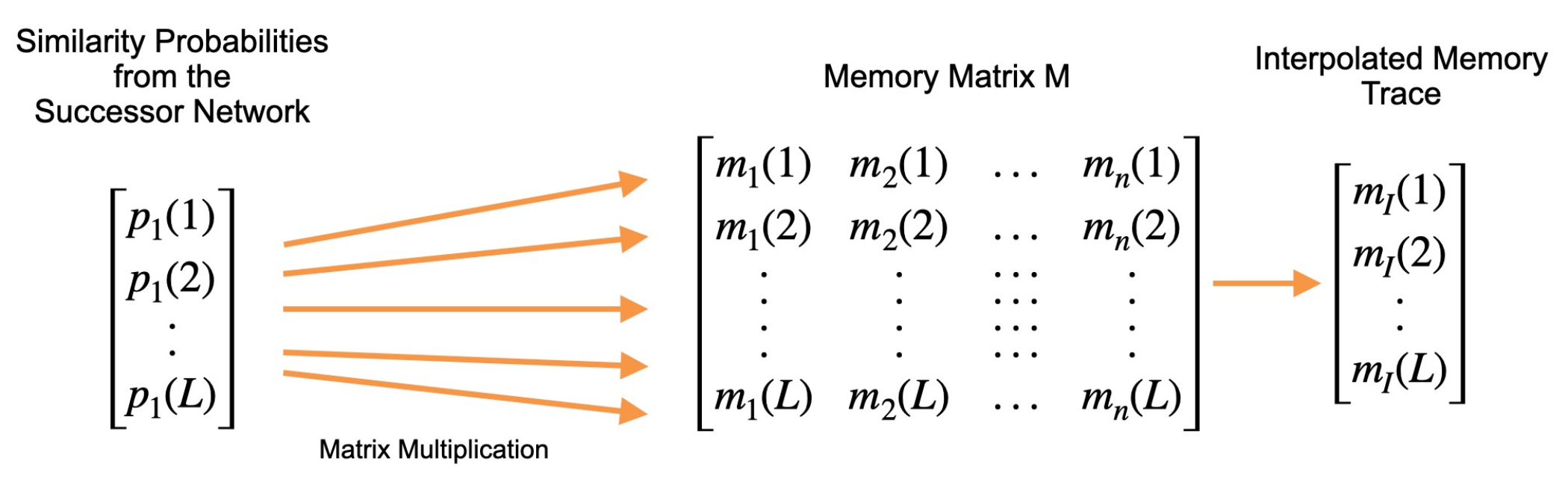}
    \caption{Schematic representation of the interpolation process. The predicted vector from the neural network $p_1$ gets multiplied with the training data, called Memory Matrix $M$, which gives the interpolated memory trace $m_I$.}
    \label{fig3}
\end{figure}

To be able to interpolate one modality from the other, we use the predicted vector from the neural network after training. The vector can be seen as a weighted pointer to the training data matrix, we call memory matrix. The pointer is used for a matrix multiplication with the memory matrix and the result is the interpolated memory trace (cf. fig. \ref{fig3}).
Therefore, we can use incomplete inputs, in our case empty images, empty word embeddings or noisy data as input for the trained network. The prediction for these cases can then be used for interpolating the inputs by using the described method.

\subsection*{Multi-dimensional scaling}
A frequently used method to generate low-dimensional embeddings of high-dimensional data is t-distributed stochastic neighbor embedding (t-SNE) \cite{van2008visualizing}. However, in t-SNE the resulting low-dimensional projections can be highly dependent on the detailed parameter settings \cite{wattenberg2016use}, sensitive to noise, and may not preserve, but rather often scramble the global structure in data \cite{vallejos2019exploring, moon2019visualizing}.
In contrast to that, multi-Dimensional-Scaling (MDS) \cite{torgerson1952multidimensional, kruskal1964nonmetric,kruskal1978multidimensional,cox2008multidimensional} is an efficient embedding technique to visualize high-dimensional point clouds by projecting them onto a 2-dimensional plane. Furthermore, MDS has the decisive advantage that it is parameter-free and all mutual distances of the points are preserved, thereby conserving both the global and local structure of the underlying data. 

When interpreting patterns as points in high-dimensional space and dissimilarities between patterns as distances between corresponding points, MDS is an elegant method to visualize high-dimensional data. By color-coding each projected data point of a data set according to its label, the representation of the data can be visualized as a set of point clusters. For instance, MDS has already been applied to visualize for instance word class distributions of different linguistic corpora \cite{schilling2021analysis}, hidden layer representations (embeddings) of artificial neural networks \cite{schilling2021quantifying,krauss2021analysis}, structure and dynamics of recurrent neural networks \cite{krauss2019analysis, krauss2019recurrence, krauss2019weight, metzner2022dynamics}, or brain activity patterns assessed during e.g. pure tone or speech perception \cite{krauss2018statistical,schilling2021analysis}, or even during sleep \cite{krauss2018analysis,traxdorf2019microstructure}. 
In all these cases the apparent compactness and mutual overlap of the point clusters permits a qualitative assessment of how well the different classes separate.

\subsection*{Generalized Discrimination Value}
We used the GDV to calculate cluster separability as published and explained in detail in \cite{schilling2021quantifying}. Briefly, we consider $N$ points $\mathbf{x_{n=1..N}}=(x_{n,1},\cdots,x_{n,D})$, distributed within $D$-dimensional space. A label $l_n$ assigns each point to one of $L$ distinct classes $C_{l=1..L}$. In order to become invariant against scaling and translation, each dimension is separately z-scored and, for later convenience, multiplied with $\frac{1}{2}$:
\begin{align}
s_{n,d}=\frac{1}{2}\cdot\frac{x_{n,d}-\mu_d}{\sigma_d}.
\end{align}
Here, $\mu_d=\frac{1}{N}\sum_{n=1}^{N}x_{n,d}\;$ denotes the mean, and $\sigma_d=\sqrt{\frac{1}{N}\sum_{n=1}^{N}(x_{n,d}-\mu_d)^2}$ the standard deviation of dimension $d$.
Based on the re-scaled data points $\mathbf{s_n}=(s_{n,1},\cdots,s_{n,D})$, we calculate the {\em mean intra-class distances} for each class $C_l$ 
\begin{align}
\bar{d}(C_l)=\frac{2}{N_l (N_l\!-\!1)}\sum_{i=1}^{N_l-1}\sum_{j=i+1}^{N_l}{d(\textbf{s}_{i}^{(l)},\textbf{s}_{j}^{(l)})},
\end{align}
and the {\em mean inter-class distances} for each pair of classes $C_l$ and $C_m$
\begin{align}
\bar{d}(C_l,C_m)=\frac{1}{N_l  N_m}\sum_{i=1}^{N_l}\sum_{j=1}^{N_m}{d(\textbf{s}_{i}^{(l)},\textbf{s}_{j}^{(m)})}.
\end{align}
Here, $N_k$ is the number of points in class $k$, and $\textbf{s}_{i}^{(k)}$ is the $i^{th}$ point of class $k$.
The quantity $d(\textbf{a},\textbf{b})$ is the euclidean distance between $\textbf{a}$ and $\textbf{b}$. Finally, the Generalized Discrimination Value (GDV) is calculated from the mean intra-class and inter-class distances  as follows:
\begin{align}
\mbox{GDV}=\frac{1}{\sqrt{D}}\left[\frac{1}{L}\sum_{l=1}^L{\bar{d}(C_l)}\;-\;\frac{2}{L(L\!-\!1)}\sum_{l=1}^{L-1}\sum_{m=l+1}^{L}\bar{d}(C_l,C_m)\right]
 \label{GDVEq}
\end{align}

\noindent whereas the factor $\frac{1}{\sqrt{D}}$ is introduced for dimensionality invariance of the GDV with $D$ as the number of dimensions.

\vspace{0.2cm}\noindent Note that the GDV is invariant with respect to a global scaling or shifting of the data (due to the z-scoring), and also invariant with respect to a permutation of the components in the $N$-dimensional data vectors (because the euclidean distance measure has this symmetry). The GDV is zero for completely overlapping, non-separated clusters, and it becomes more negative as the separation increases. A GDV of -1 signifies already a very strong separation.

\subsection*{Code Implementation}
The models were coded in Python 3.10. The neural networks were design using the Keras \cite{keras} library with TensorFlow \cite{tensorflow2015-whitepaper}. Mathematical operations were performed with numpy \cite{numpy}, scikit-learn \cite{scikit-learn} and the spaCy \cite{spacy2} libraries.
Visualizations were realized with matplotlib \cite{matplot}.

\section*{Results}

\subsection*{Learning the Cognitive Map}
\begin{figure}[h]
\centering
\includegraphics[width=1.0\linewidth]{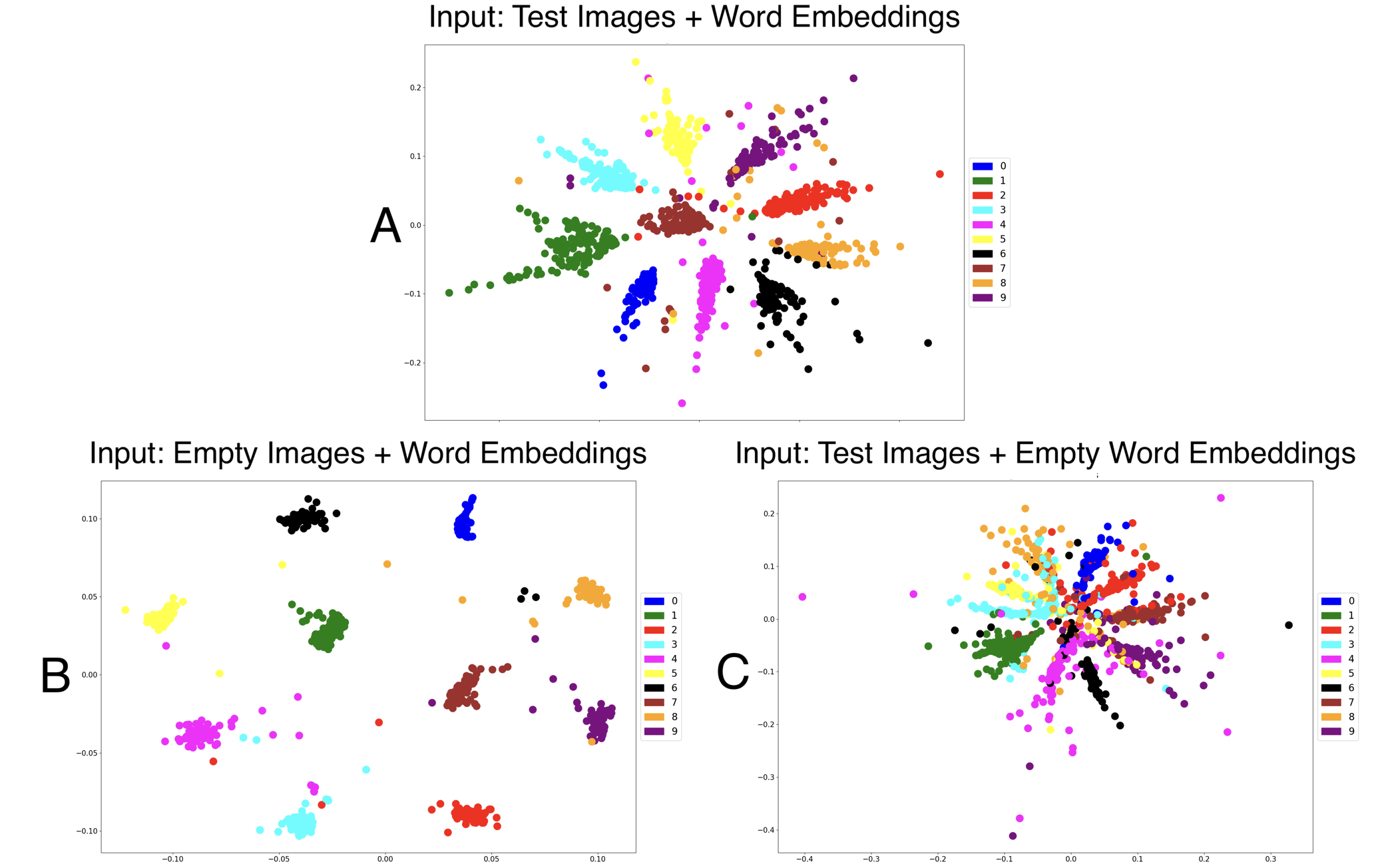}
\caption{Visualization of Predictive Patterns by Trained Neural Networks on Different Input Modalities. Neural networks trained with 1,000 images are used to generate predictions across various input types, which are then represented as Multidimensional Scaling (MDS) graphs. A: MDS representation for combined input modalities, illustrating distinct clusters corresponding to each digit. B: Predictions based solely on word embeddings, showing clustering for each digit. Unlike other modalities, these clusters are not arranged in a circular order but are more dispersed within the feature space. C: Using only test images as input, the resulting clusters resemble those in A, but are positioned more closely to one another.}
\label{fig4}
\end{figure}


We find that the neural networks are adept at learning representations of cognitive maps. This observation is based on a network that underwent training and testing with a set of 1,000 images and word embeddings. Through the use of Multidimensional Scaling (MDS) graphs generated from the test data, we examine the network's performance in three different scenarios: processing both input modalities together, and processing either images or word embeddings independently. In each of these cases, we qualitatively observe the formation of distinct clusters in the network's predictions for the ten digits (cf. Figure \ref{fig4}). In order to quantify the degree of clustering, we also calculated the Generalized Discrimination Value (GDV, cf. Methods) for all three graphs. In the case where the network receives both modalities, we end up with a GDV of $-0.1233$ (cf. Figure \ref{fig4}A). However, if the images as input modality are missing, we find a GDV of $-0.1158$ (cf. Figure \ref{fig4}B). In the last scenario where the word embeddings are missing, we find a GDV of $-0.1281$ (cf. Figure \ref{fig4}C). Hence, all GDVs are close to each other and indicate strong clustering.

\begin{figure}[h]
\centering
\includegraphics[width=1.0\linewidth]{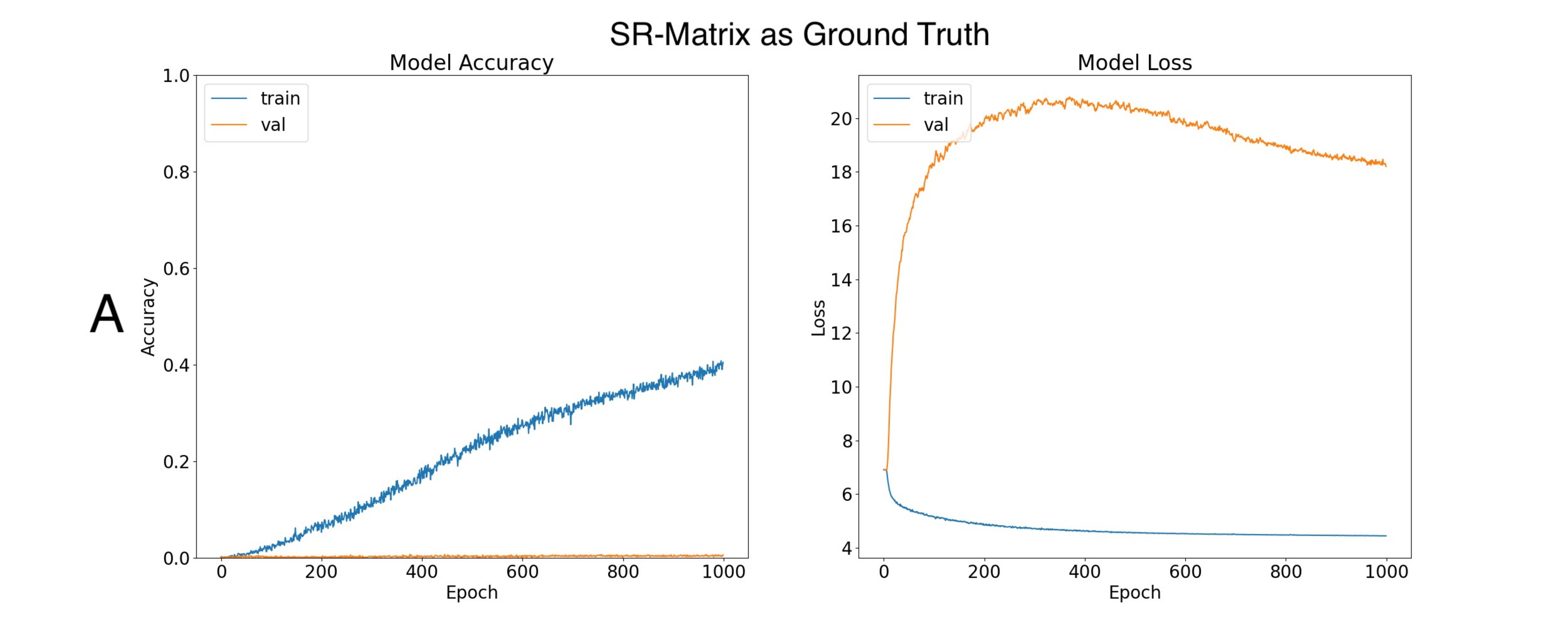}\\
\includegraphics[width=1.0\linewidth]{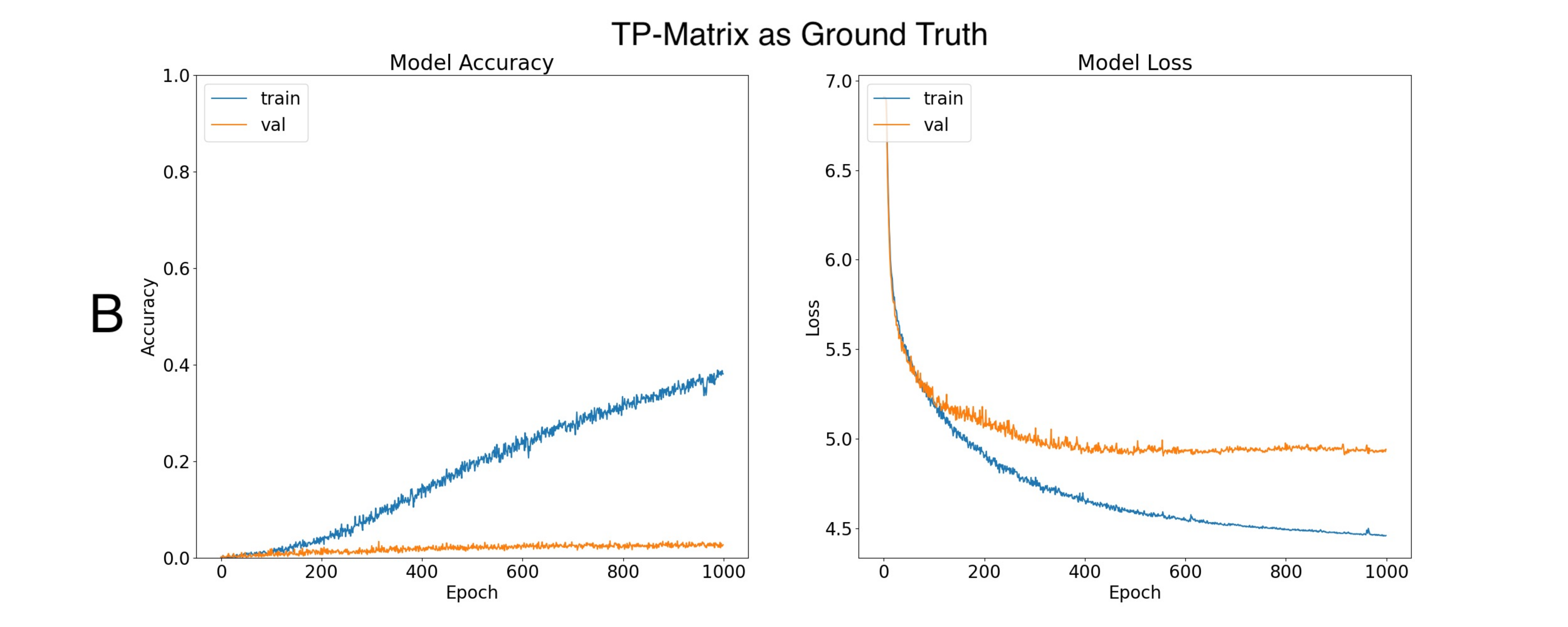}
\caption{Training Performance of a Network with 1,000 Training Images. A: Training with the SR-Matrix as the label shows promising accuracy, reaching approximately 40\% after 1,000 epochs, alongside a rapid decrease in loss. However, the validation accuracy remains near 0\%, and the loss initially increases before plateauing. B: Utilizing the TP-Matrix as the training and testing label yields varying outcomes. The training accuracy and loss trends are similar to those observed in A. Notably, the validation accuracy climbs to about 5\%, and the loss shows a continuous decline.}
\label{fig5}
\end{figure}

The model accuracy and loss is also shown for the network which used 1000 images for training and testing (cf. Figure \ref{fig5}). Again, we test two different scenarios. First, with the successor representation (SP) matrix set as ground truth for testing (cf. Figure \ref{fig5}A), and second, with the state transition probability (TP) matrix set as ground truth (cf. Figure \ref{fig5}B).

In the present study, we observed a consistent increase in training accuracy, culminating at approximately 40\% after 1000 epochs. Concurrently, the training loss exhibited a rapid decline, stabilizing post approximately 400 epochs. Extended training duration further enhanced the training accuracy, surpassing 90\% subsequent to 5000 epochs. Utilizing the SR-Matrix as the ground truth, the validation accuracy remained marginal, near 0\%, with the loss exhibiting a nominal reduction towards the end. In contrast, employing the TP-Matrix as the label, the validation accuracy attained a level of approximately 5\%, alongside a steady diminution in loss. It is pertinent to reiterate that the labels utilized in both the training and testing phases were not one-hot-encoded class labels, but rather transition probability vectors.

\newpage
\subsection*{Inference of Different Modalities}

\begin{figure}[h]
\centering
\includegraphics[width=0.9\linewidth]{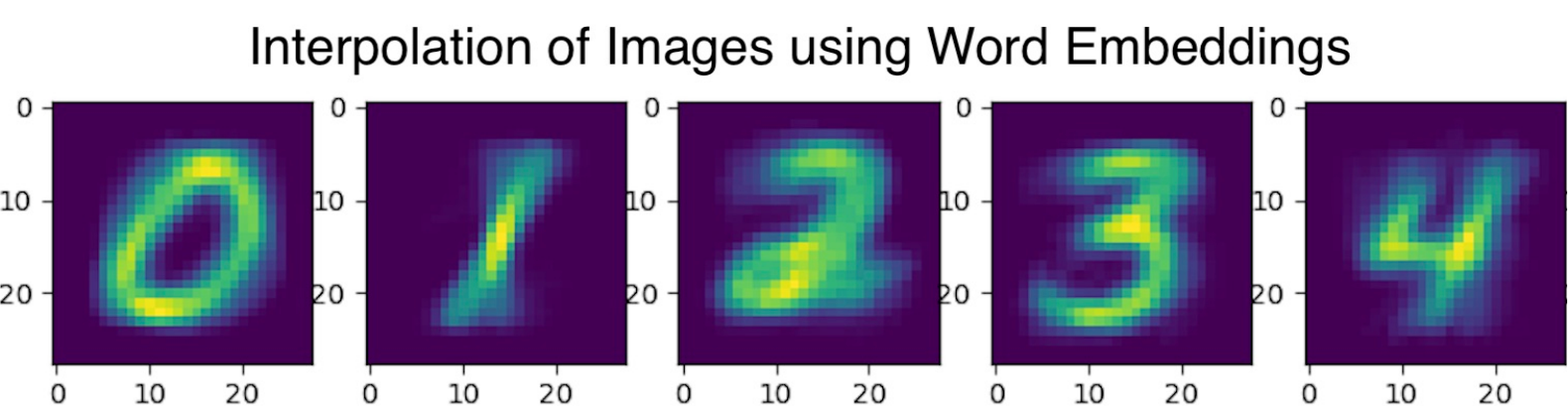}\\
\includegraphics[width=0.9\linewidth]{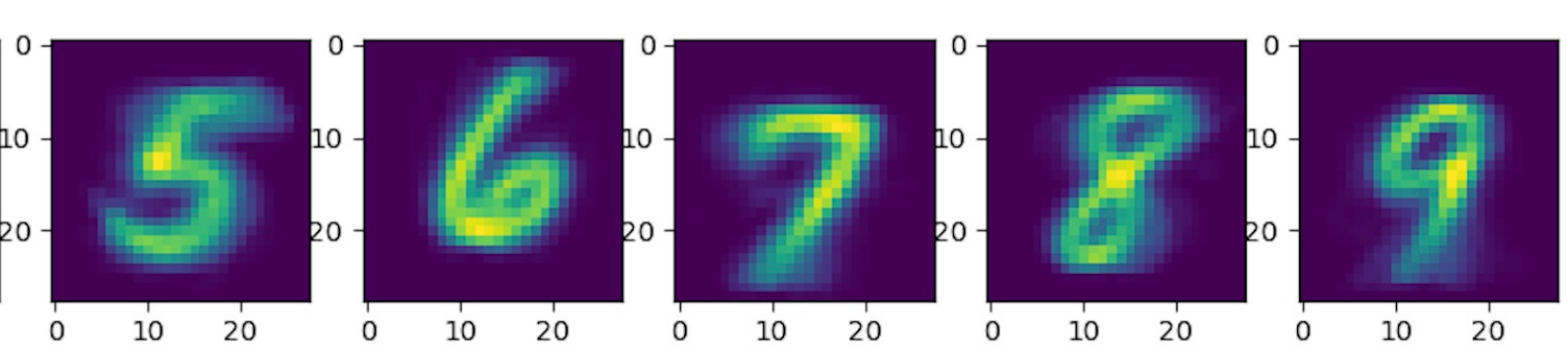}
\caption{Inference of images using solely word embeddings as inputs for the neural network: The network generates a predicted vector, which is subsequently employed for inference with either the training or memory matrix. Predictions are systematically arranged from 0 to 9 corresponding to the word embeddings. Qualitative analysis reveals distinct image representations that accurately correspond to each input embedding for the respective numbers.}
\label{fig6}
\end{figure}

In the subsequent phase, the pre-trained network was employed to facilitate cross-modality inference. Initially, the focus was on image inference utilizing noisy word embeddings as inputs. The qualitative assessment of the interpolation results revealed that the network accurately inferred all digits ranging from 0 to 9. Each number was visually discernible and corresponded precisely with the respective word embedding (cf. Figure \ref{fig6}).

\begin{figure}[h]
\centering
\includegraphics[width=0.9\linewidth]{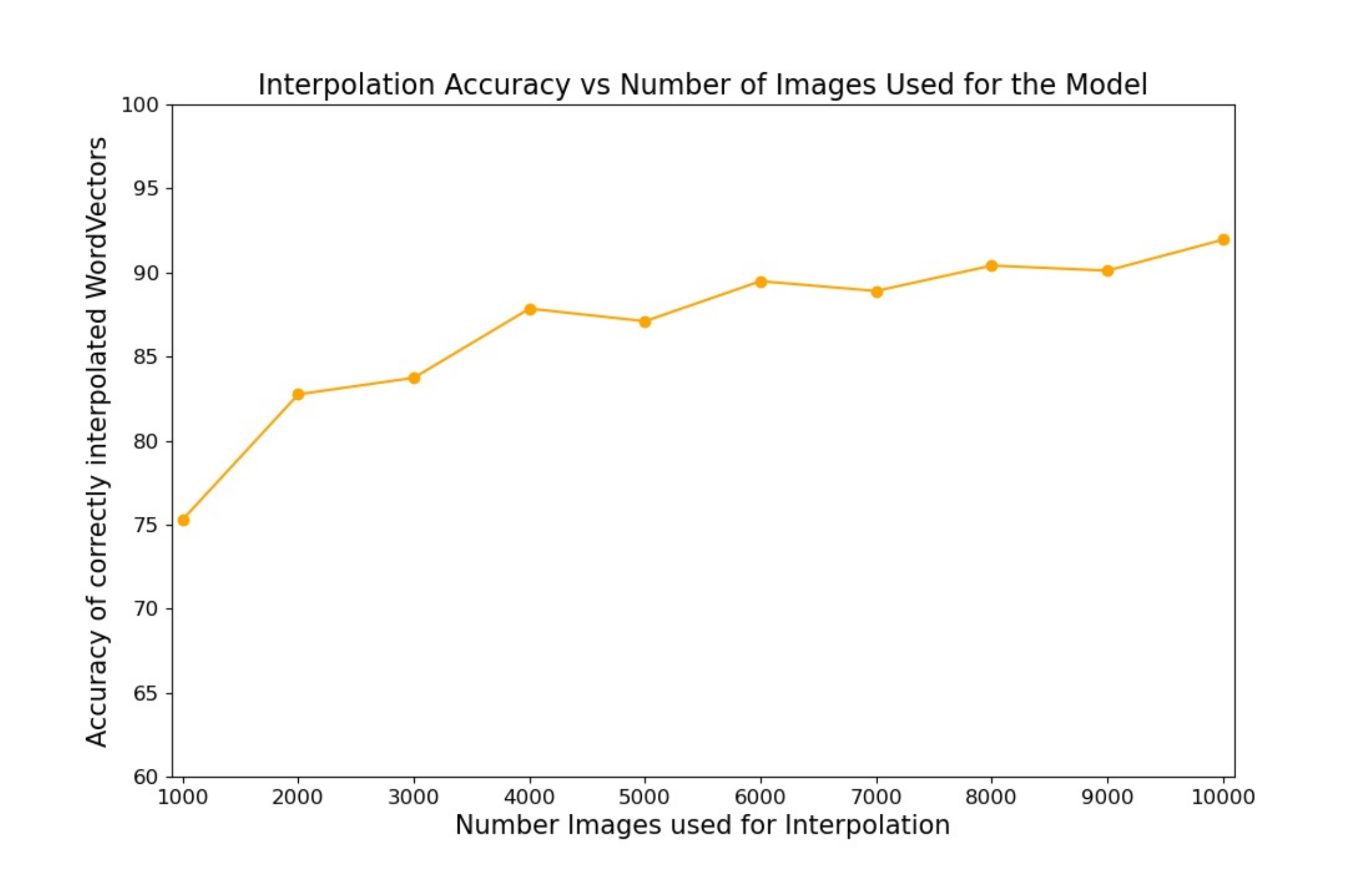}\\
\caption{Using test images for inferring corresponding word embeddings: The network is provided with these test images and initially empty word embeddings to make predictions. Subsequently, the predicted pointers are applied to infer the embeddings. The graph juxtaposes the number of images utilized in network training against the accuracy of correctly inferring word embeddings. It is observed that accuracy increases in tandem with the number of images used, commencing at 75\% accuracy with 1,000 images and reaching over 92\% with 10,000 images.}
\label{fig7}
\end{figure}

In the subsequent analysis, the test images were employed to infer the corresponding word embeddings. The accuracy of this inference varied based on the number of images utilized during the training phase, which were subsequently used for inference. The results indicate a positive correlation between the number of images and inference accuracy. Specifically, employing 1,000 images for inference yielded an accuracy of 75\%. This accuracy progressively increased with the quantity of images used, culminating at 93\% when the number of images reached 10,000 (cf. Figure \ref{fig7}).

\newpage
\section*{Discussion}

In this study, our development of a multi-modal cognitive map has revealed its beneficial properties. The designed network adeptly learns to recognize similarities between various states and their respective features. A notable achievement of this approach is the inference accuracy of the word embeddings, which exceeds 90\%. Furthermore, the qualitative analysis of image inference indicates a clear and accurate representation for each number. This progress marks a significant extension from our previous work on cognitive maps, which primarily focused on handcrafted feature states \cite{stoewer2023neural}. By integrating semantic word embeddings and visual representations, we have expanded the scope and application of cognitive maps. Additionally, the incorporation of a feature extraction component for image state representation and a refined preprocessing technique for the word embedding feature vector has further enhanced the efficacy of our approach.

The difficulty encountered in accurately learning the correct successor representation label for the test data can be attributed to the matrix multiplication method used in generating the SR-Matrix from the TP-Matrix. This multiplication process results in each state being highly likely to be its own successor, a characteristic that presents a challenge for test states. Without the ability to identify their position within the matrix, these test states are unable to accurately learn the successor matrix, leading them instead to learn general transition probabilities. Additionally, our efforts to replicate the variably scaled maps observed in previous research were unsuccessful. These maps, which demonstrate varying scales along the entorhinal cortex as discussed in \cite{collin2015memory}, remain an objective for future investigations into multi-modal maps. This highlights an area for ongoing research, underscoring the potential for further advancements in understanding and replicating complex neural mapping processes.

This study highlights the efficacy of cognitive maps as a conceptual tool. Despite utilizing a dataset from a relatively narrow domain, our findings suggest the potential for integrating diverse modalities—such as images, feature representations, spatial locations, and time—within a cognitive map framework. A key strength of our approach lies in the ability to customize the memory base or experiential foundation for the model, allowing for controlled learning and interpolation processes. This aspect could be particularly beneficial in emerging fields like Large Language Models (LLMs). LLMs have seen remarkable improvements in recent years, especially in text generation capabilities. However, as noted by Mahowald et al. \cite{mahowald2023}, these models often exhibit formal competence without functional competence. Given that LLMs are predominantly trained on word sequences and lack a grounded understanding of how these words relate to real-world contexts, integrating a cognitive map could provide valuable contextual information. By enriching word embeddings with visual elements, factual data, and spatial or temporal dimensions, cognitive maps could enhance the functional understanding of LLMs, bridging the gap between textual data and its real-world associations.

The concept of cognitive maps, which are epitomized in the mammalian brain by the entorhinal-hippocampal complex, serves as an effective framework for organizing and recalling information from a vast array of memories. This biological paradigm presents a fascinating model that could be mirrored in the development of Large Language Models (LLMs). The potential for LLMs to either emulate or incorporate cognitive map structures offers a promising research direction. In the biological model, the hippocampus is central to information processing across a network of interconnected cortical areas. Similarly, LLMs might adopt cognitive maps as a fundamental mechanism for backend processing. This integration could significantly enhance their ability to process and retrieve contextual information. Such an advancement in LLMs would likely lead to a substantial improvement in their ability to comprehend and remember complex contexts, resulting in interactions that are more intuitive and akin to human-like understanding.

The grounding problem in artificial intelligence has been a pivotal issue, focusing on how to connect symbolic or abstract representations to real-world entities and experiences. The incorporation of multi-modal inputs in the neural network approach we discuss offers a potential solution to this challenge. By integrating diverse data forms, such as images and word embeddings, Large Language Models (LLMs) gain access to a diverse array of information. This integration enables the bridging of gaps between different types of data, fostering a network where various modalities are interlinked. Such a setup facilitates a level of context-awareness previously unattainable, allowing LLMs to comprehend and ground abstract concepts through the associations learned across these modalities. Consequently, in scenarios involving novel or incomplete information, the LLM could leverage its cognitive map to deduce missing context, thereby demonstrating an advanced ability to ground abstract concepts in tangible, real-world experiences.

Future research endeavors will be directed towards integrating a broader range of modalities to create a more comprehensive representation of the environment. Advanced architectures such as transformers offer promising avenues for enhancing the model's comprehension of state spaces. Transformers have already been explored in the context of graph networks, as highlighted in \cite{NEURIPS2019_graphs}. Given that the cognitive map we designed resembles a weighted directed graph, the successor representation could potentially benefit from the application of transformer architectures. The transformer's proficiency in the next word prediction task, where it essentially learns to point to known vocabulary elements, aligns well with the methodology of our approach. Adapting the transformer to our model could facilitate a deeper understanding of more intricate and diverse input types, leveraging its attention mechanism to handle larger and more complex data sets effectively.

Advancements in the field of computational neuroscience, particularly in comprehending the hippocampal formation and modeling place and grid cells, could contribute significantly to the enhancement of current deep learning models.

In the realm of artificial intelligence, effective planning remains a challenging area. Two decades ago, Schmidhuber et al. introduced concepts for hierarchical planning architectures, as outlined in their foundational works \cite{Schmidhuber1990TowardsCL,Schmidhuber1991LearningTG, schmidhuber_1993}. These early concepts have continued to evolve, with recent developments further enriching the field \cite{schmidhuber2015learning, schmidhuber2018big}. An integral component of advancing AI planning capabilities lies in the comprehensive modeling of grid and place cell dynamics, particularly considering the variation in grid sizes. Such a model holds the potential to facilitate more adaptable and flexible planning strategies for the future. The ability to adjust for different planning horizons could enable AI systems to project far into the future or focus on immediate upcoming events, enhancing their decision-making and predictive capacities.

Conversely, models inspired by neurology have the potential to enhance our comprehension of existing deep learning models. Contemporary concerns and apprehensions regarding the future of AI systems, as discussed in recent literature \cite{metz2023the, yudkowsky2023pausing}, are partly rooted in a limited understanding of their internal mechanisms \cite{liang2021explaining,frank_baby_2023} and the inherent biases within these systems \cite{Ntoutsi2020,schwartz2022towards,munn_uselessness_2023}. By integrating insights from neurological models, there is an opportunity to demystify the inner workings of deep learning models, addressing the issues of transparency and bias, which are critical for the responsible development and deployment of future AI technologies. This approach could lead to more ethically aligned and comprehensible AI systems, mitigating the fears associated with their advancement.

Neuroscience based method could also help here to advance AI alignment research to find known brain structures and find similar behavior in neural networks. Our study proposes a foundational methodology for learning from environmental interactions and the interrelationships between objects within that environment. By doing so, it facilitates a nuanced and realistic comprehension of the world's context. Moreover, it enhances our understanding of memory processing, which is pivotal for deciphering the workings of underlying neural networks. In an era dominated by increasingly large and complex systems, such as extensive Language Models (LLMs), this understanding is crucial for enhancing transparency in their result generation processes. Additionally, this study holds potential applications in AI Alignment. By gaining greater control over the storage and processing of memories, we can more effectively guide the behavior of neural networks, ensuring their alignment with human values and enhancing their safety. Therefore, computational neuroscience could not only advance AI systems but lead a road to more transparent and saver system in general.

\FloatBarrier
\section*{Availability of Data and Materials}
The datasets used and/or analysed during the current study are available from the corresponding author on reasonable request.

\section*{Acknowledgments}

This work was funded by the Deutsche Forschungsgemeinschaft (DFG, German Research Foundation): grants KR\,5148/2-1 (project number 436456810), KR\,5148/3-1 (project number 510395418) and GRK\,2839 (project number 468527017) to PK, and grant SCHI\,1482/3-1 (project number 451810794) to AS.

\section*{Author contributions}
PS performed computer simulations and prepared all figures. PS, AM and PK designed the study. PK and AM supervised the study. PS, AS, AM and PK discussed the results and wrote the manuscript.

\section*{Competing interests}
The authors declare no competing financial interests.

\FloatBarrier
\bibliographystyle{unsrt}
\bibliography{literature}

\begin{thebibliography}{10}

\bibitem{mahowald2023}
Kyle Mahowald, Anna Ivanova, Idan Blank, Nancy Kanwisher, Joshua Tenenbaum, and
  Evelina Fedorenko.
\newblock Dissociating language and thought in large language models: a
  cognitive perspective.
\newblock {\em arxiv}, 2023.

\bibitem{marcus2018deep}
Gary Marcus.
\newblock Deep learning: A critical appraisal.
\newblock {\em arXiv preprint arXiv:1801.00631}, 2018.

\bibitem{opitz_memory_2014}
Bertram Opitz.
\newblock Memory function and the hippocampus.
\newblock {\em Frontiers of Neurology and Neuroscience}, 34:51--59, 2014.

\bibitem{epstein2017cognitive}
Russell~A Epstein, Eva~Zita Patai, Joshua~B Julian, and Hugo~J Spiers.
\newblock The cognitive map in humans: spatial navigation and beyond.
\newblock {\em Nature neuroscience}, 20(11):1504--1513, 2017.

\bibitem{park2021inferences}
Seongmin~A Park, Douglas~S Miller, and Erie~D Boorman.
\newblock Inferences on a multidimensional social hierarchy use a grid-like
  code.
\newblock {\em bioRxiv}, pages 2020--05, 2021.

\bibitem{killian_grid_2018}
Nathaniel~J. Killian and Elizabeth~A. Buffalo.
\newblock Grid cells map the visual world.
\newblock {\em Nature Neuroscience}, 21(2), 2018.

\bibitem{rolls2015limbic}
Edmund~T Rolls.
\newblock Limbic systems for emotion and for memory, but no single limbic
  system.
\newblock {\em cortex}, 62:119--157, 2015.

\bibitem{o1971hippocampus}
John O'Keefe and Jonathan Dostrovsky.
\newblock The hippocampus as a spatial map: preliminary evidence from unit
  activity in the freely-moving rat.
\newblock {\em Brain research}, 1971.

\bibitem{hafting2005microstructure}
Torkel Hafting, Marianne Fyhn, Sturla Molden, May-Britt Moser, and Edvard~I
  Moser.
\newblock Microstructure of a spatial map in the entorhinal cortex.
\newblock {\em Nature}, 436(7052):801--806, 2005.

\bibitem{derdikman_fragmentation_2009}
Dori Derdikman, Jonathan~R Whitlock, Albert Tsao, Marianne Fyhn, Torkel
  Hafting, May-Britt Moser, and Edvard~I Moser.
\newblock Fragmentation of grid cell maps in a multicompartment environment.
\newblock {\em Nature Neuroscience}, 12(10):1325--1332, October 2009.

\bibitem{o1978hippocampus}
John O'keefe and Lynn Nadel.
\newblock {\em The hippocampus as a cognitive map}.
\newblock Oxford university press, 1978.

\bibitem{moser2017spatial}
Edvard~I Moser, May-Britt Moser, and Bruce~L McNaughton.
\newblock Spatial representation in the hippocampal formation: a history.
\newblock {\em Nature neuroscience}, 20(11):1448--1464, 2017.

\bibitem{bellmund2018navigating}
Jacob~LS Bellmund, Peter G{\"a}rdenfors, Edvard~I Moser, and Christian~F
  Doeller.
\newblock Navigating cognition: Spatial codes for human thinking.
\newblock {\em Science}, 362(6415), 2018.

\bibitem{collin2015memory}
Silvy~HP Collin, Branka Milivojevic, and Christian~F Doeller.
\newblock Memory hierarchies map onto the hippocampal long axis in humans.
\newblock {\em Nature neuroscience}, 18(11):1562--1564, 2015.

\bibitem{whittington2020tolman}
James~CR Whittington, Timothy~H Muller, Shirley Mark, Guifen Chen, Caswell
  Barry, Neil Burgess, and Timothy~EJ Behrens.
\newblock The tolman-eichenbaum machine: Unifying space and relational memory
  through generalization in the hippocampal formation.
\newblock {\em Cell}, 183(5):1249--1263, 2020.

\bibitem{stachenfeld2014design}
Kimberly~L Stachenfeld, Matthew Botvinick, and Samuel~J Gershman.
\newblock Design principles of the hippocampal cognitive map.
\newblock {\em Advances in neural information processing systems},
  27:2528--2536, 2014.

\bibitem{stachenfeld2017hippocampus}
Kimberly~L Stachenfeld, Matthew~M Botvinick, and Samuel~J Gershman.
\newblock The hippocampus as a predictive map.
\newblock {\em Nature neuroscience}, 20(11):1643, 2017.

\bibitem{stoewer_neural_2022}
Paul Stoewer, Christian Schlieker, Achim Schilling, Claus Metzner, Andreas
  Maier, and Patrick Krauss.
\newblock Neural network based successor representations to form cognitive maps
  of space and language.
\newblock {\em Scientific Reports}, 12:11233, 2022.
\newblock Number: 1 Publisher: Nature Publishing Group.

\bibitem{stoewer2023neural}
Paul Stoewer, Achim Schilling, Andreas Maier, and Patrick Krauss.
\newblock Neural network based formation of cognitive maps of semantic spaces
  and the putative emergence of abstract concepts.
\newblock {\em Scientific Reports}, 13(1):3644, 2023.

\bibitem{surendra2023word}
Kishore Surendra, Achim Schilling, Paul Stoewer, Andreas Maier, and Patrick
  Krauss.
\newblock Word class representations spontaneously emerge in a deep neural
  network trained on next word prediction.
\newblock {\em arXiv preprint arXiv:2302.07588}, 2023.

\bibitem{stoewer2023conceptual}
Paul Stoewer, Achim Schilling, Andreas Maier, and Patrick Krauss.
\newblock Conceptual cognitive maps formation with neural successor networks
  and word embeddings, 2023.

\bibitem{SR_Original}
Peter Dayan.
\newblock {Improving Generalization for Temporal Difference Learning: The
  Successor Representation}.
\newblock {\em Neural Computation}, 5(4):613--624, 07 1993.

\bibitem{deng2012mnist}
Li~Deng.
\newblock The mnist database of handwritten digit images for machine learning
  research.
\newblock {\em IEEE Signal Processing Magazine}, 29(6):141--142, 2012.

\bibitem{spacy2}
Matthew Honnibal and Ines Montani.
\newblock {spaCy 2}: Natural language understanding with {B}loom embeddings,
  convolutional neural networks and incremental parsing.
\newblock To appear, 2017.

\bibitem{van2008visualizing}
Laurens Van~der Maaten and Geoffrey Hinton.
\newblock Visualizing data using t-sne.
\newblock {\em Journal of machine learning research}, 9(11), 2008.

\bibitem{wattenberg2016use}
Martin Wattenberg, Fernanda Vi{\'e}gas, and Ian Johnson.
\newblock How to use t-sne effectively.
\newblock {\em Distill}, 1(10):e2, 2016.

\bibitem{vallejos2019exploring}
Catalina~A Vallejos.
\newblock Exploring a world of a thousand dimensions.
\newblock {\em Nature biotechnology}, 37(12):1423--1424, 2019.

\bibitem{moon2019visualizing}
Kevin~R Moon, David van Dijk, Zheng Wang, Scott Gigante, Daniel~B Burkhardt,
  William~S Chen, Kristina Yim, Antonia van~den Elzen, Matthew~J Hirn, Ronald~R
  Coifman, et~al.
\newblock Visualizing structure and transitions in high-dimensional biological
  data.
\newblock {\em Nature biotechnology}, 37(12):1482--1492, 2019.

\bibitem{torgerson1952multidimensional}
Warren~S Torgerson.
\newblock Multidimensional scaling: I. theory and method.
\newblock {\em Psychometrika}, 17(4):401--419, 1952.

\bibitem{kruskal1964nonmetric}
Joseph~B Kruskal.
\newblock Nonmetric multidimensional scaling: a numerical method.
\newblock {\em Psychometrika}, 29(2):115--129, 1964.

\bibitem{kruskal1978multidimensional}
Joseph~B Kruskal.
\newblock {\em Multidimensional scaling}.
\newblock Number~11. Sage, 1978.

\bibitem{cox2008multidimensional}
Michael~AA Cox and Trevor~F Cox.
\newblock Multidimensional scaling.
\newblock In {\em Handbook of data visualization}, pages 315--347. Springer,
  2008.

\bibitem{schilling2021analysis}
Achim Schilling, Rosario Tomasello, Malte~R Henningsen-Schomers, Alexandra
  Zankl, Kishore Surendra, Martin Haller, Valerie Karl, Peter Uhrig, Andreas
  Maier, and Patrick Krauss.
\newblock Analysis of continuous neuronal activity evoked by natural speech
  with computational corpus linguistics methods.
\newblock {\em Language, Cognition and Neuroscience}, 36(2):167--186, 2021.

\bibitem{schilling2021quantifying}
Achim Schilling, Andreas Maier, Richard Gerum, Claus Metzner, and Patrick
  Krauss.
\newblock Quantifying the separability of data classes in neural networks.
\newblock {\em Neural Networks}, 139:278--293, 2021.

\bibitem{krauss2021analysis}
Patrick Krauss, Claus Metzner, Nidhi Joshi, Holger Schulze, Maximilian
  Traxdorf, Andreas Maier, and Achim Schilling.
\newblock Analysis and visualization of sleep stages based on deep neural
  networks.
\newblock {\em Neurobiology of sleep and circadian rhythms}, 10:100064, 2021.

\bibitem{krauss2019analysis}
Patrick Krauss, Alexandra Zankl, Achim Schilling, Holger Schulze, and Claus
  Metzner.
\newblock Analysis of structure and dynamics in three-neuron motifs.
\newblock {\em Frontiers in Computational Neuroscience}, 13:5, 2019.

\bibitem{krauss2019recurrence}
Patrick Krauss, Karin Prebeck, Achim Schilling, and Claus Metzner.
\newblock Recurrence resonance” in three-neuron motifs.
\newblock {\em Frontiers in computational neuroscience}, page~64, 2019.

\bibitem{krauss2019weight}
Patrick Krauss, Marc Schuster, Verena Dietrich, Achim Schilling, Holger
  Schulze, and Claus Metzner.
\newblock Weight statistics controls dynamics in recurrent neural networks.
\newblock {\em PloS one}, 14(4):e0214541, 2019.

\bibitem{metzner2022dynamics}
Claus Metzner and Patrick Krauss.
\newblock Dynamics and information import in recurrent neural networks.
\newblock {\em Frontiers in Computational Neuroscience}, 16, 2022.

\bibitem{krauss2018statistical}
Patrick Krauss, Claus Metzner, Achim Schilling, Konstantin Tziridis, Maximilian
  Traxdorf, Andreas Wollbrink, Stefan Rampp, Christo Pantev, and Holger
  Schulze.
\newblock A statistical method for analyzing and comparing spatiotemporal
  cortical activation patterns.
\newblock {\em Scientific reports}, 8(1):1--9, 2018.

\bibitem{krauss2018analysis}
Patrick Krauss, Achim Schilling, Judith Bauer, Konstantin Tziridis, Claus
  Metzner, Holger Schulze, and Maximilian Traxdorf.
\newblock Analysis of multichannel eeg patterns during human sleep: a novel
  approach.
\newblock {\em Frontiers in human neuroscience}, 12:121, 2018.

\bibitem{traxdorf2019microstructure}
Maximilian Traxdorf, Patrick Krauss, Achim Schilling, Holger Schulze, and
  Konstantin Tziridis.
\newblock Microstructure of cortical activity during sleep reflects respiratory
  events and state of daytime vigilance.
\newblock {\em Somnologie}, 23(2):72--79, 2019.

\bibitem{keras}
Fran\c{c}ois Chollet et~al.
\newblock Keras, 2015.

\bibitem{tensorflow2015-whitepaper}
Mart\'{\i}n Abadi, Ashish Agarwal, Paul Barham, Eugene Brevdo, Zhifeng Chen,
  Craig Citro, Greg~S. Corrado, Andy Davis, Jeffrey Dean, Matthieu Devin,
  Sanjay Ghemawat, Ian Goodfellow, Andrew Harp, Geoffrey Irving, Michael Isard,
  Yangqing Jia, Rafal Jozefowicz, Lukasz Kaiser, Manjunath Kudlur, Josh
  Levenberg, Dandelion Man\'{e}, Rajat Monga, Sherry Moore, Derek Murray, Chris
  Olah, Mike Schuster, Jonathon Shlens, Benoit Steiner, Ilya Sutskever, Kunal
  Talwar, Paul Tucker, Vincent Vanhoucke, Vijay Vasudevan, Fernanda Vi\'{e}gas,
  Oriol Vinyals, Pete Warden, Martin Wattenberg, Martin Wicke, Yuan Yu, and
  Xiaoqiang Zheng.
\newblock {TensorFlow}: Large-scale machine learning on heterogeneous systems,
  2015.
\newblock Software available from tensorflow.org.

\bibitem{numpy}
Charles~R. Harris, K.~Jarrod Millman, St{'{e}}fan~J. van~der Walt, Ralf
  Gommers, Pauli Virtanen, David Cournapeau, Eric Wieser, Julian Taylor,
  Sebastian Berg, Nathaniel~J. Smith, Robert Kern, Matti Picus, Stephan Hoyer,
  Marten~H. van Kerkwijk, Matthew Brett, Allan Haldane, Jaime~Fern{'{a}}ndez
  del R{'{\i}}o, Mark Wiebe, Pearu Peterson, Pierre G{'{e}}rard-Marchant, Kevin
  Sheppard, Tyler Reddy, Warren Weckesser, Hameer Abbasi, Christoph Gohlke, and
  Travis~E. Oliphant.
\newblock Array programming with {NumPy}.
\newblock {\em Nature}, 585(7825):357--362, September 2020.

\bibitem{scikit-learn}
F.~Pedregosa, G.~Varoquaux, A.~Gramfort, V.~Michel, B.~Thirion, O.~Grisel,
  M.~Blondel, P.~Prettenhofer, R.~Weiss, V.~Dubourg, J.~Vanderplas, A.~Passos,
  D.~Cournapeau, M.~Brucher, M.~Perrot, and E.~Duchesnay.
\newblock Scikit-learn: Machine learning in {P}ython.
\newblock {\em Journal of Machine Learning Research}, 12:2825--2830, 2011.

\bibitem{matplot}
J.~D. Hunter.
\newblock Matplotlib: A 2d graphics environment.
\newblock {\em Computing in Science \& Engineering}, 9(3):90--95, 2007.

\bibitem{NEURIPS2019_graphs}
Seongjun Yun, Minbyul Jeong, Raehyun Kim, Jaewoo Kang, and Hyunwoo~J Kim.
\newblock Graph transformer networks.
\newblock In H.~Wallach, H.~Larochelle, A.~Beygelzimer, F.~d\textquotesingle
  Alch\'{e}-Buc, E.~Fox, and R.~Garnett, editors, {\em Advances in Neural
  Information Processing Systems}, volume~32. Curran Associates, Inc., 2019.

\bibitem{Schmidhuber1990TowardsCL}
Jurgen Schmidhuber.
\newblock Towards compositional learning in dynamic networkstechnical report.
\newblock 1990.

\bibitem{Schmidhuber1991LearningTG}
Jurgen Schmidhuber.
\newblock Learning to generate subgoals for action sequences.
\newblock {\em IJCNN-91-Seattle International Joint Conference on Neural
  Networks}, ii:453 vol.2--, 1991.

\bibitem{schmidhuber_1993}
J\"{u}rgen Schmidhuber and Reiner Wahnsiedler.
\newblock Planning simple trajectories using neural subgoal generators.
\newblock In {\em Proceedings of the Second International Conference on From
  Animals to Animats 2: Simulation of Adaptive Behavior: Simulation of Adaptive
  Behavior}, page 196–202, Cambridge, MA, USA, 1993. MIT Press.

\bibitem{schmidhuber2015learning}
Juergen Schmidhuber.
\newblock On learning to think: Algorithmic information theory for novel
  combinations of reinforcement learning controllers and recurrent neural world
  models, 2015.

\bibitem{schmidhuber2018big}
Juergen Schmidhuber.
\newblock One big net for everything, 2018.

\bibitem{metz2023the}
Cade Metz.
\newblock The godfather of ai leaves google and warns of danger ahead.
\newblock {\em New York Times}, 2023.

\bibitem{yudkowsky2023pausing}
Eliezer Yudkowsky.
\newblock Pausing ai developments isn't enough. we need to shut it all down.
\newblock {\em Time Magazine (Mar, 2023). https://time.
  com/6266923/ai-eliezer-yudkowsky-open-letter-not-enough}, 2023.

\bibitem{liang2021explaining}
Yu~Liang, Siguang Li, Chungang Yan, Maozhen Li, and Changjun Jiang.
\newblock Explaining the black-box model: A survey of local interpretation
  methods for deep neural networks.
\newblock {\em Neurocomputing}, 419:168--182, 2021.

\bibitem{frank_baby_2023}
Michael~C. Frank.
\newblock Baby steps in evaluating the capacities of large language models.
\newblock {\em Nature Reviews Psychology}, June 2023.

\bibitem{Ntoutsi2020}
Eirini Ntoutsi, Pavlos Fafalios, Ujwal Gadiraju, Vasileios Iosifidis, Wolfgang
  Nejdl, Maria-Esther Vidal, Salvatore Ruggieri, Franco Turini, Symeon
  Papadopoulos, Emmanouil Krasanakis, Ioannis Kompatsiaris, Katharina
  Kinder-Kurlanda, Claudia Wagner, Fariba Karimi, Miriam Fernandez, Harith
  Alani, Bettina Berendt, Tina Kruegel, Christian Heinze, Klaus Broelemann,
  Gjergji Kasneci, Thanassis Tiropanis, and Steffen Staab.
\newblock Bias in data-driven artificial intelligence systems---an introductory
  survey.
\newblock {\em WIREs Data Mining and Knowledge Discovery}, 10(3):e1356, 2020.

\bibitem{schwartz2022towards}
Reva Schwartz, Apostol Vassilev, Kristen Greene, Lori Perine, Andrew Burt,
  Patrick Hall, et~al.
\newblock Towards a standard for identifying and managing bias in artificial
  intelligence.
\newblock {\em NIST special publication}, 1270(10.6028), 2022.

\bibitem{munn_uselessness_2023}
Luke Munn.
\newblock The uselessness of {AI} ethics.
\newblock {\em AI and Ethics}, 3(3):869--877, August 2023.

\end{thebibliography}

\end{document}